\documentclass[twocolumn]{article}

\usepackage{amsmath,siunitx}
\usepackage{graphicx}
\usepackage{multirow}
\usepackage{booktabs}
\usepackage{float}
\usepackage{subfig}
\usepackage{authblk}

\graphicspath{{./Images}}

\title{Model-based Deep Learning\\for High-Dimensional Periodic Structures}

\author[1]{Lucas Polo-López}
\author[1]{Luc Le Magoarou}
\author[2]{Romain~Contreres}
\author[1]{María García-Vigueras}
\affil[1]{Institut d'Electronique et des Technologies du numéRique (IETR), UMR CNRS 6164, INSA Rennes, France}
\affil[2]{Antenna Department, Centre National d'Études Spatiales (CNES), Toulouse, France}
\date{}

\begin{document}

\maketitle

\begin{abstract}
This work presents a deep learning surrogate model for the fast simulation of high-dimensional frequency selective surfaces. We consider unit-cells which are built as multiple concatenated stacks of screens and their design requires the control over many geometrical degrees of freedom. Thanks to the introduction of physical insight into the model, it can produce accurate predictions of the S-parameters of a certain structure after training with a reduced dataset.
The proposed model is highly versatile and it can be used with any kind of frequency selective surface, based on either perforations or patches of any arbitrary geometry. Numeric examples are presented here for the case of frequency selective surfaces composed of screens with rectangular perforations, showing an excellent agreement between the predicted performance and such obtained with a full-wave simulator.

\end{abstract}

\section{Introduction}
Artificial intelligence is currently revolutionising the way information is processed. In particular, deep learning (DL) techniques have proven to be powerful tools to embrace the modelling of high-dimensional datasets in a very efficient way \cite{Goodfellow2016, Lecun2015}. These techniques have enabled the treatment of scenarios of great complexity such as natural language processing \cite{Brown2020}, image generation \cite{Goodfellow2020} or protein folding prediction \cite{Jumper2021}, just to mention a few.

An emerging trend in DL consists in using \textit{a priori} knowledge of the problem under study to help overcoming the main drawbacks of training a DL model (e.g. requirement of massive data sets, computational burden of training...). This is referred to in the literature as \textit{model-based DL} \cite{Shlezinger2021,Yang2020,Zappone2019}. It can be seen as a way to make existing models data-adaptive, and has been applied with success to several 
 aspects of the physical layer of wireless communication systems \cite{He2019}, including MIMO channel estimation \cite{Yassine2022} and detection \cite{Samuel2017}.

The design of radio-frequency (RF) structures can also benefit significantly from the use of DL, specially when considering architectures of high complexity (e.g. time-reconfigurable three-dimensional structures with high number of degrees of freedom). Illustrative examples are emerging recently showing that the assistance from DL can be key for the successful design of reflectarrays and metasurfaces \cite{Prado2018, Naseri2020, Lin2020, Goudos2022,Fallah2023}. 

Classically, these structures have been developed based on equivalent models. In particular, different solutions are available these days that allow to drive the design based on physical insight on the device electromagnetic (EM) behavior \cite{Costa2012,Medina2010,Mesa2015,Conde-Pumpido2022}. The complexity of such models may increase drastically when the geometry of the unit cell presents a high number of degrees of freedom. This leads to a drastic increment on the time required for their development. Such is the case of frequency selective surfaces (FSS) based on multiple stacking. In \cite{Mesa2018} an approach for their efficient characterization by means of an equivalent circuit is proposed. However, this approach limits the structure degrees of freedom, since it can only consider screens of equal geometry.

\begin{figure}[bt]
\centering
\includegraphics[width=\columnwidth]{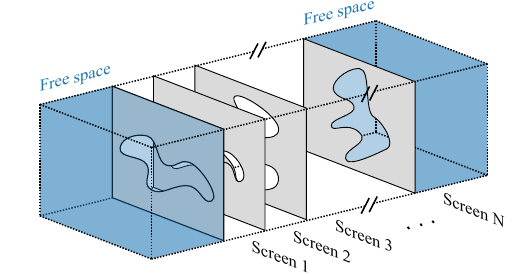}
\caption{Unit cell of a multi-stacked FSS with perforated screens of arbitrary geometry.}
\label{fig:Unit_Cell_3D}
\end{figure}

The goal of this work is to illustrate the great potential of using model-based DL for the characterization of high-dimensional periodic structures (which can be used as FSS, polarizers, field-shaping radomes, phased arrays...). More specifically, we consider unit cells as the one illustrated in Fig.~\ref{fig:Unit_Cell_3D}, composed of multiple stacks of perforated screens with a high number of geometrical degrees of freedom. We propose to combine DL with prior knowledge about the physical phenomena taking place in such type of RF structure, and we build a surrogate model (SM). For a certain FSS geometry and a set of frequency points, the goal of this SM is to predict associated S-parameters without the need of making a full-wave (FW) simulation. SMs can be coarsely divided in two groups: function approximation models, which approximate samples corresponding to FW simulation results \cite{Kim2007}, and physical-based models, which are based on some kind of lower-fidelity representation of the EM phenomena involved in the problem. The later group, to which the SM presented in this work belongs, typically show better generalization capabilities than function approximation models \cite{Koziel2017,Koziel2023}.

\section{Surrogate Model for FSS}

\begin{figure}[t]
\centering
\includegraphics[]{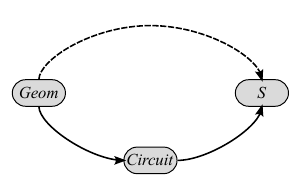}
\caption{Different approaches to create a DL SM for characterizing a FSS. The dashed line depicts a model that directly relates the geometry ($Geom$) of the FSS to its scattering parameters ($S$), without making use of \emph{a priori} knowledge. The solid line represents a SM that makes use of knowledge about the EM problem under study to establish an equivalent circuit as an intermediate step in the calculation of $S$.}
\label{fig:Alternative_Paradigms}
\end{figure}

When facing the problem of creating a DL SM that characterizes a FSS, the first idea that comes to mind is trying to establish a direct relationship between the geometry of the FSS and its scattering parameters. This approach is depicted in Fig.~\ref{fig:Alternative_Paradigms} by the dashed arrow, and although it has been applied with success in the past \cite{Kim2007}, it requires an elevated amount of computational resources. Hence, it will be extremely costly (or directly impossible) to employ this kind of approach for establishing a SM of a high-dimensional FSS.

Alternatively, it is possible to benefit from \emph{a priori} knowledge of the electromagnetic problem under study to establish a SM with a higher computational efficiency \cite{Koziel2017,Koziel2023}. For the scenario under study of modelling a FSS, an approach like that depicted by the solid arrow in Fig.~\ref{fig:Alternative_Paradigms} can be followed. It is possible to establish an equivalent circuit that models the frequency response of the FSS with good accuracy. Given a set of values for the circuit parameters, the S-parameters of the circuit (and consequently those of the corresponding FSS) can be obtained with a negligible computational cost. Therefore, instead of developing a DL model that relates the geometry of the FSS to its frequency response, it is possible to develop a DL model that establishes a relationship between the geometry of the FSS and its equivalent circuit. Since the number of parameters to predict will be much smaller (only a few circuit element values, instead of several samples of the S-parameters over the frequency band of interest), the complexity and therefore the computational burden of training such SM will be significantly smaller.

\begin{figure}[t]
\centering
\includegraphics[width=\columnwidth]{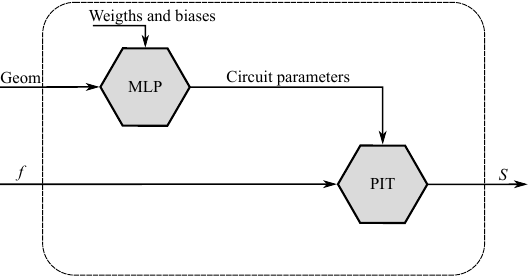}
\caption{Block diagram of the proposed surrogate model (SM) with physical insight.}
\label{fig:Surrogate_Model_Schema}
\end{figure}

A block diagram of the proposed SM is depicted in Fig.~\ref{fig:Surrogate_Model_Schema}. The model presents two inputs, a \emph{geometry} (Geom) of the FSS and a vector of frequency points ($f$). In this work the word geometry is used to refer to the perforation shapes and the separation between screens of a certain FSS realization. A classical Multilayer Perceptron (MLP) with Rectified Linear Unit (ReLU) activations \cite{Glorot2011} is used to make a prediction of the circuit parameters of an equivalent circuit corresponding to the input geometry. The topology of this circuit is defined in the Physical Insight Topology (PIT) module which is in charge of, given a certain set of circuit parameters, calculate the S-parameters ($S$, in Fig.~\ref{fig:Surrogate_Model_Schema}) of that circuit at a certain input frequency.

It is important to highlight that the PIT is tailored for the specific FSS for which the SM is being developed. Each screen is represented by a certain number of lumped elements, and these sets of lumped elements are cascaded by transmission lines representing the physical separation between the screens (the electric permittivity and magnetic permeability of the line are adjusted accordingly to the dielectric between the screens). The more precisely the lumped elements characterize each of the screens, the more accurate the S-parameter predictions will be. Put in other way, \emph{the more physical insight is put into the SM}, the more accurate the results will be.

Once that the circuit topology is defined, the PIT can be implemented by computing the ABCD-matrix of each circuit element, multiplying them all to obtain the global matrix of the circuit and then transforming the result to its equivalent S-matrix \cite{Frickey1994}. This process is computationally very efficient since only basic algebraic operations are involved.

It should be noted that all of the above is valid for any multi-stacked FSS regardless of the number of screens and the geometry of their perforations. Moreover, despite the discussion has been focused on FSSs composed of perforated screens, this concept can also be applied to FSS composed by a stacking of patches (or even a combination of patches and perforations) just by implementing the adequate PIT.

\subsection{PIT for perforated screens}

In order to produce numerical examples, for the rest of the paper the PIT will be particularized for a stacking of conductive screens with rectangular perforations (or slots) of width \SI{1}{\milli\metre}. The length of the slots is a design parameter (degree of freedom), along with the distance between each pair of screens, to define the geometry of the FSS. Additionally, the material between the screens is considered as vacuum, and the periodicity of the FSS is \SI{18}{\milli\metre} both in the horizontal and vertical dimensions.

\begin{figure*}[t]
\centering
\includegraphics[width=\textwidth]{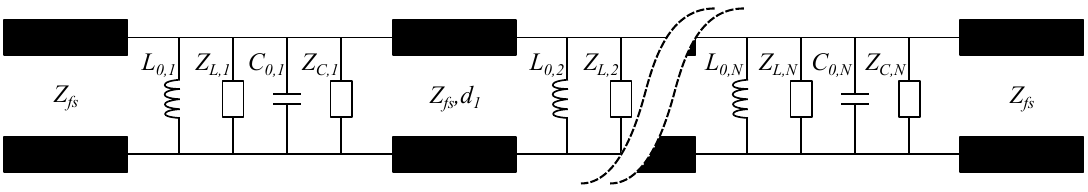}
\caption{Equivalent circuit of a multistacked FSS composed of N perforated screens.}
\label{fig:Equivalent_Circuit}
\vspace{-\baselineskip}
\end{figure*}

Under these circumstances, the $i$-th screen of the FSS can be characterized by the following equivalent admittance:
\begin{align}
    Y_{eq,i}(\omega) =\frac{1}{j\omega L_{0,i}}&+\sum_{k=1}^{N_{\mathrm{TE}}}\alpha_{L,i}Y_{\mathrm{TE},k}(\omega) +\nonumber\\
    + j\omega C_{i,0} &+ \sum_{k=1}^{N_{\mathrm{TM}}}\alpha_{C,i}Y_{\mathrm{TM},k}(\omega),\label{eq:Y_eq}
\end{align}
where the summations characterize the first $N_{\mathrm{TE}}$ and $N_{\mathrm{TM}}$ Floquet harmonics with the lowest cutoff frequencies, while the highest order harmonics are taken into account by $L_{0,i}$ and $C_{0,i}$ \cite{Garcia-Vigueras2012}. Since only normal incidence will be considered in the examples, a single high order Floquet harmonic will be considered explicitly and therefore $N_{\mathrm{TE}}=N_{\mathrm{TM}}=1$, which yields the circuit depicted in Fig.~\ref{fig:Equivalent_Circuit} for the case of a FSS with N perforated screens. However, it must be emphasized that the complexity of the PIT could be increased to include more harmonics if necessary.

\section{Surrogate Model Training}

The dataset necessary for training the SM can be generated by means of a FW simulation. A parametric sweep can be performed over the design parameters to obtain the simulated S-parameters for different geometries. Specifically, in this work CST Microwave Studio has been used for such task.

The training process has been performed by means of an implementation of the Adam algorithm \cite{Kingma2017} (in particular, that included in the PyTorch library \cite{Paszke2019}). The weights and biases of the MLP are initialized at random; however, for the case of the PIT better convergence is obtained if more \textit{a priori} knowledge is introduced into the SM by performing an initial guess of the circuit parameters. For each geometry in the training dataset, each of its screens is simulated independently using the FW simulator. Then, a numeric fitting of the parameters in \eqref{eq:Y_eq} can be done to adjust the results obtained by this partial simulation. Of course, this approach completely neglects the interactions between the screens, but the idea is that this procedure allows to obtain a relatively accurate initial guess of the circuit parameters without a significant computational burden (since a single screen is simulated at each time, these simulations take little time to run).

\begin{figure}[t]
\centering
\includegraphics[]{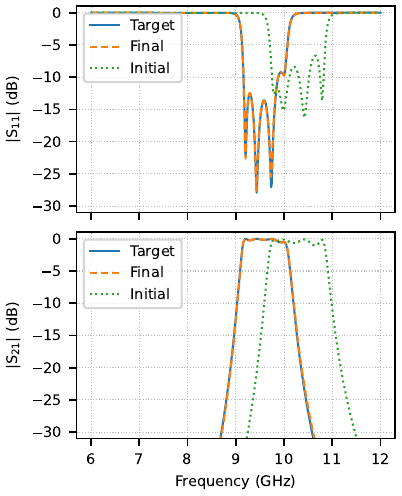}
\caption{Scattering parameters provided by the PIT for the circuit parameters in Table~\ref{tab:FSS_Analysis} compared with the FW simulation (Target).}\label{fig:FSS_Analysis_Sparams}
\end{figure}

To illustrate the validity of this approach, a numeric example of the PIT training is presented. In this case a FSS of $\mathrm{N}=4$ screens is considered. The length of each of the slots is $l_{slot,i}=\{14.91, 14.80, 14.75, 14.88\}\:\si{\milli\metre}$, and the separations between them are $d_{i}=\{10.3, 8.79, 10.3\}\:\si{\milli\metre}$. Table~\ref{tab:FSS_Analysis} presents both the initial guess of the circuit parameters and also the final values after the training process. It can be observed that their variation is quite significant, arriving to more than 20\% for several of them. Nevertheless, as illustrated by Fig.~\ref{fig:FSS_Analysis_Sparams}, they provide a good departure point which allows to obtain a result that perfectly matches the target value.

{
%
\setlength{\tabcolsep}{2mm}%
\renewcommand{\arraystretch}{1.2}
\newcommand{\CPcolumnonewidth}{not used}%
\newcommand{\CPcolumntwowidth}{21mm}%
\newcommand{\CPcolumnthreewidth}{12mm}%
\newcommand{\CPcolumnfourwidth}{33mm}%
\begin{table}[t]
\caption{Circuit parameters before and after the training of the PIT for a certain geometry of a $\mathrm{N}=4$ FSS.}
\small
\centering
    \begin{tabular}{cccc}
    \toprule
        \textbf{Parameter} & \textbf{Initial} & \textbf{Final} & \textbf{Variation (\%)}\\\midrule
        $L_{0,1}\:(\si{\henry})$ & 1.6129e-09 & 1.8522e-09 & +14.83 \\
        $C_{0,1}\:(\si{\farad})$ & 2.7936e-13 & 2.9064e-13 & +4.04 \\
        $\alpha_{L1}$ & 2.2065e-07 & 2.4792e-07 & +12.36 \\
        $\alpha_{C1}$ & 4.1445e-11 & 4.2703e-11 & +3.04 \\
        $L_{0,2}\:(\si{\henry})$ & 1.4629e-09 & 1.5292e-09 & +4.53 \\
        $C_{0,2}\:(\si{\farad})$ & 2.8419e-13 & 3.5589e-13 & +25.23 \\
        $\alpha_{L2}$ & 2.1672e-07 & 2.0562e-07 & -5.12 \\
        $\alpha_{C2}$ & 4.1393e-11 & 4.9911e-11 & +20.58 \\
        $L_{0,3}\:(\si{\henry})$ & 1.3657e-09 & 1.4612e-09 & +6.99 \\
        $C_{0,3}\:(\si{\farad})$ & 2.8153e-13 & 3.7264e-13 & +32.36 \\
        $\alpha_{L3}$ & 2.1577e-07 & 2.0361e-07 & -5.64 \\
        $\alpha_{C3}$ & 4.1400e-11 & 5.2064e-11 & +25.76 \\
        $L_{0,4}\:(\si{\henry})$ & 1.6215e-09 & 1.7276e-09 & +6.54 \\
        $C_{0,4}\:(\si{\farad})$ & 2.7925e-13 & 3.4225e-13 & +22.56 \\
        $\alpha_{L4}$ & 2.1763e-07 & 2.1389e-07 & -1.72 \\
        $\alpha_{C4}$ & 4.1329e-11 & 4.8811e-11 & +18.10 \\
        $d_{1}\:(\si{\milli\metre})$ & 10.3000 & 10.2788 & -0.21 \\
        $d_{2}\:(\si{\milli\metre})$ & 8.7904 & 8.2938 & -5.65 \\
        $d_{3}\:(\si{\milli\metre})$ & 10.3000 & 9.9791 & -3.12 \\\bottomrule
    \end{tabular}
\label{tab:FSS_Analysis}
\end{table}
}

It should be noted that the presented training approach requires to train the PIT independently prior to the MLP training. This is necessary for the PIT to be able to benefit from the \textit{a priori} knowledge about the circuit parameters. In a future work, it would be possible to perform a \emph{fine tuning} of the surrogate model parameters by including an additional training step were both models are trained simultaneously.

\section{Results}
To illustrate the capabilities of the proposed SM, a numeric example of FSS with $\mathrm{N} = 2$ is presented in this section. The periodicity of the FSS is \SI{18}{\milli\metre} and the width of the slots is defined as \SI{1}{\milli\metre}. There are therefore three design parameters that will define the FSS geometry: the length of each of the slots and the distance between the two screens. The dataset is generated by varying the length of the slots between \SI{9.5}{\milli\metre} and \SI{15}{\milli\metre} (9 samples) and the distance between the screens from \SI{7}{\milli\metre} to \SI{15}{\milli\metre} (9 samples). This yields a total of $N_{sample} = 729$ samples, and for each of them a FW simulation has been performed to obtain its corresponding S-parameters at a given set of $N_{freq} = 200$ frequencies. This process produces the following labeled dataset:
\begin{equation}
   \Big\{\mathbf{g}_i,\big\{S_{21,i}^{goal}(f_j)\big\}_{j=1}^{N_{freq}}\Big\}_{i=1}^{N_{sample}},
\end{equation}\label{eq:Dataset}
\noindent where $\mathbf{g}_i$ is a vector containing the geometric parameters of the structure for the $i$-th sample and $S_{21,i}^{goal}(f_j)$ is the associated S-parameter at the $j$-th frequency.
It should be noted that the size of this dataset is relatively moderate for a DL problem. However, as it is shown below, thanks to the aforementioned introduction of physical insight the SM achieves accurate predictions even with such a reduced dataset.

The samples are then randomized and 80\% of them are assigned to the train dataset and the other 20\% are reserved for the test dataset. The DL model allows to predict S-parameters at each considered frequency as a function of an input geometry. For the $i$-th input and $j$-th frequency, the output is denoted $S_{21}^{model}(\mathbf{g}_i,f_j)$. The SM is trained using the Adam algorithm mentioned earlier along with the following cost function:
\begin{equation}
    Cost=\sum^{N_{samp}}_i\sum^{N_{freq}}_j \frac{|S_{21}^{model}(\mathbf{g}_i,f_j)-S_{21,i}^{goal}(f_j)|}{N_{samp}N_{freq}},
\end{equation}\label{eq:Cost_Function}
\indent After the training, the SM is evaluated over the test dataset to assess its performance. The cost values resulting from the different samples present a mean of 0.036 and a standard deviation of 0.017.

\begin{figure}[!h]
\centering
\includegraphics[]{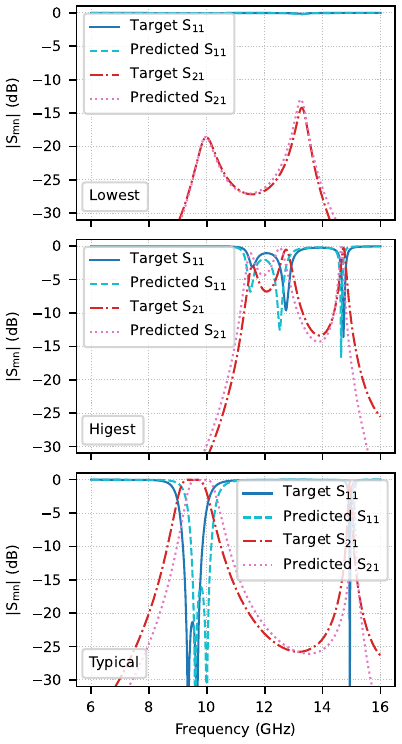}
\caption{Scattering parameters predicted by the SM with the FW simulation for the test cases with the lowest and highest cost value, and also for a typical case.}\label{fig:Surrogate_SParam_All_Mag}
\end{figure}

Figure~\ref{fig:Surrogate_SParam_All_Mag} presents the S-parameters produced by the SM for the cases with the lowest and highest cost value. As it can be observed, even for the case with the highest cost value the agreement between the curves produced by the SM and a FW simulator (labeled as Target) is reasonably good.

It might seem that the response of these two cases is strange in the sense that it does not present the characteristic filtering response of a FSS. However, it should be remarked that the training and tests datasets have been generated by a systematic parametric sweep of the FSS to model. Therefore, many of the samples do not present a combination of the design parameters that yields an actual filtering response. Nevertheless, it should be remarked that the objective of the SM is to behave as similarly as possible to the FW simulator regardless of the input geometry.

Anyway, to demonstrate that the SM can be effectively used as a design tool, Fig.~\ref{fig:Surrogate_SParam_All_Mag} also presents the case of a geometry that was neither in the training nor in the test datasets which has been designed to present a pass-band response (labeled as Typical). As it can be seen, in this case the SM also arrives to estimate the S-parameters of the structure with good accuracy.

\section{Conclusion}

This work has presented a model-based deep learning (DL) surrogate model (SM) for multi-stacked frequency selective surfaces (FSS), which is capable of predicting the S-parameters of the FSS without the need of making a full-wave (FW) simulation.

The SM is composed of a Multilayer Perceptron (MLP) that works in combination with a Physical Insight Topology (PIT) module. The inclusion of physical insight into the model allows to obtain accurate predictions with a reduced training dataset.

Although the numerical examples presented in this work considered a FSS with linear slots, this does not reduce the generality of the proposed SM, which can be applied to FSSs with any geometry. Moreover, the good results obtained in this initial work show the great potential of application that model-based DL has for the characterization and design of periodic structures with complex geometries and many degrees of freedom.

\section*{Acknowledgements}
This publication is supported by the french region of Brittany, Ministry of Higher Education and Research, Rennes Métropole and Conseil Départemental 35, through under grant SAD volet 1-2019-3Debris and also by the Centre National d'Études Spatiales (CNES) under contract R\&T RS21/TC-0007-144.

\bibliographystyle{ieeetr}
\bibliography{Model_Based_DL}

\end{document}